\documentclass[preprint,5p,times]{elsarticle}
\usepackage[version=4]{mhchem}
\usepackage{graphics}
\usepackage{graphicx}
\usepackage{float}
\usepackage{xcolor}
\usepackage{amssymb}
\usepackage{tabularx}

\newcommand{\fg}[2]{\langle \langle #1 ; #2 \rangle \rangle}
\newcommand{\vek}{\vec{k}} 

\newcommand{\veq}{\vec{k}+\vec{Q}}
\newcommand{\kaq}{\vec{k}+\vec{Q}}
\newcommand{\ka}{\vec{k}}

\newcommand\red[1]{{\color{red}#1}}

\usepackage{amssymb}
\usepackage{amsmath}

\begin{document}

\begin{frontmatter}



\title{Interplay between the charge density wave phase and a pseudogap under antiferromagnetic correlations}


\author[ufrgs]{L. C. Prauchner} 
\author[ufsm]{E. J. Calegari}
\author[andres]{J. Faundez}
\author[ufrgs]{S. G. Magalhaes}
\address[ufrgs]{Instituto de Física, Universidade Federal do Rio Grande do Sul, 91501-970, Porto Alegre, RS, Brazil}
\address[ufsm]{Departamento de F\'{\i}sica, Universidade Federal de Santa Maria, 97105-900, Santa Maria, RS, Brazil}
\address[andres]{Departamento de Ciencias Físicas, Universidad Andres Bello, Santiago 837-0136, Chile}

\begin{abstract}
In this study, we explore the impact of short-range antiferromagnetic correlations on the charge density wave (CDW) phase in strongly correlated electron systems exhibiting the pseudogap phenomenon. Our investigation employs an n-pole approximation to consider the repulsive Coulomb interaction $(U)$ and antiferromagnetic correlations. Considering an one-band Hubbard model to account for the Coulomb interaction and a BCS-like model for the CDW order parameter, we observed that an increase in $U$ enhances antiferromagnetic fluctuations, resulting in a flattened re-normalized band around the antinodal point $(\pi,0)$. The pseudogap manifests itself in the band structure and density of states, prompting exploration of various $U$ and occupation number values. Our findings indicate that antiferromagnetic correlations significantly influence the CDW state, as the Fermi surface is reconstructed within the ordered phase. Furthermore, we found a Lifshitz transition inside both the CDW phase and the normal state, with the latter preceding the onset of the pseudogap.
\end{abstract}




\end{frontmatter}



\section{Introduction}
\label{sec1}

The pseudogap is a very enigmatic and yet not fully understood phenomenon. Characterized by suppression of states around the Fermi level below a certain temperature $T^{*}$ \cite{Chen2022, Bejas2011},
this phenomenon is generally observed in strongly correlated electron systems, such as cuprates \cite{keimer2015} and pnictides \cite{Li2016}. However, we can also highlight selenium-based compounds, in particular  1$T$-TaSe$_2$ \cite{Bovet2004}, which is capable of showing both a pseudogap and a charge density wave (CDW) \cite{Singh2017}.
In some cases, these compounds also have a superconducting order \cite{Chen2018, Hu2021, Levy2016} and display a phase diagram similar to that of cuprates \cite{keimer2015}. Additionally, we can mention infinite-layered nickelates \cite{Zhang20201,Zhang2021}, in which the pseudogap phenomenon is accompanied by the emergence of superconductivity, and eventually, a CDW.

A plethora of mechanisms have been proposed to explain the nature of the pseudogap phenomenon \cite{ Fauque2006, krien2022explaining,bounoua2022hidden, punk2015quantum, sakai2016hidden,Kumar,Kumar1}. In cuprates, the short-range antiferromagnetic correlations resulting from the vicinity of the antiferromagnetic phase in the underdoped regime can be of utmost importance to clarify the pseudogap phenomenon \cite{Choiniere,Naqib,Chan,Morinari,valerii,Kuzmin,Thomas}. However, 
it is important to note that the significant role played by the short-range antiferromagnetic correlations may not be exclusive to cuprates. Experimental evidence indicates that in iron pnictides, the emergence of a pseudogap is intertwined with antiferromagnetic fluctuations \cite{Moon}. Furthermore, in nickelates, the  short-range antiferromagnetic correlations may be related with a CDW and superconductivity \cite{Rossi}, or even with a pseudogap \cite{klett}.
Thus, a question arises:  how may short-range antiferromagnetic correlations  influence an interplay between a CDW and the overall pseudogap phenomenon? 

As a proposal to answer this question, we start from a BCS-like mean field theory within the Green's function equation of motion formalism to describe a CDW instability. In order to introduce the short-range antiferromagnetic correlations which are the source of the pseudogap in the present scenario, we follow the methodology considered in references \cite{sampaio,Tifrea2003,moca,calegari2016,RODRIGUEZNUNEZ200188} and  replace the normal state uncorrelated Green's function by a correlated one. The normal state correlated Green's function is obtained through an n-pole approximation \cite{Roth,beenen,calegari2005} applied to the one-band Hubbard model \cite{Hubbard}. This is a way to consider short-range antiferromagnetic correlations in the normal state. The n-pole approximation was proposed as a correction to the Hubbard-I approximation \cite{Hubbard}, which is unable to capture magnetic solutions and antiferromagnetic correlations. The limitation of the Hubbard I approximation lies in the fact that it neglects important quantities such as the spin-spin $\langle \vec{S}_i  \cdot \vec{S}_j\rangle$ and the double occupation $\langle N_i  N_j \rangle$ correlation functions.  In the n-pole approximation, these correlations which are present in the band shift $Y_{\vec{k},\sigma}$, are preserved.

In the present theory, it is worth highlighting the importance of  next nearest neighbor hopping $t_1$. It plays a fundamental role along with the total occupation number $n_T = \langle n_{\sigma}\rangle + \langle n_{-\sigma}\rangle$ and the Coulomb interaction $U$, affecting the band shift $Y_{\vec{k},\sigma}$ mainly through the spin-spin correlation function  $\langle \vec{S}_i\cdot\vec{S}_j\rangle$, which is directly related to antiferromagnetic correlations \cite{beenen}. As an effect of the antiferromagnetic correlations, the region of the quasiparticle band around the nodal point $(\pi,\pi)$ is  shifted to lower energies. In addition, the flattening related to a van Hove singularity  in $(\pi,0)$, is enhanced. These two combined effects favor
the opening of a pseudogap in the density of states (DOS), as well as at the antinodal points of the Fermi surface. Moreover, due to the enhancement of the  flattening at $(\pi,0)$, the DOS close to the van Hove singularity increases and allows the availability of a large number of electrons, favoring pair formation and stabilizing the CDW below a critical temperature $T_C$, as occurs in superconducting systems \cite{sampaio,Kauppila,Markiewicz}.
 
 \par For this work, we chose a square lattice and set the CDW order form factor as an unconventional \textit{d-}wave symmetry, in contrast with the usual \textit{s-}wave gap symmetry that has been explored in some cuprates. In cuprates \cite{Atkinson2018,Tu2019} and transition metal dichalcogenides \cite{gao2020}, the CDW coexists with other competing orders, such as superconductivity. This choice of symmetry is not arbitrary, as both \textit{s-}wave and \textit{d-}wave cases have been observed \cite{Davis2013, Mcmahon2020} and can even compete, making it difficult to determine the predominant form factor \cite{fujita2014, blackburn2013x}. However, in the present work we only investigated $d-$wave symmetry. The present theory does not necessarily assume a phononic mechanism for electron-hole pair formation. Indeed, several authors have suggested that strongly correlated electron systems may have different origins for the phase transition from the normal state to CDW
\cite{Bogdan2d, Chen2015, EHeumen}. 

As pointed out in reference \cite{sampaio}, the n-pole approximation presents some shortcomings related to the evaluation of the correlation functions present in the band shift $Y_{\vec{k},\sigma}$. However, in reference \cite{beenen}, there was a good agreement between the quasiparticle bands obtained with the n-pole approximation \cite{Roth} and those  from quantum Monte Carlo calculations \cite{Bulut}. Furthermore, it was recently show \cite{Haurie}, that although the n-pole approximation presents a solution that may violate the
Pauli principle, it provides Fermi surfaces expected for strongly correlated
materials. 

The methodology used here, which  starts with the Green's functions from a BCS-like mean field theory and replaces the normal state uncorrelated Green's function with a correlated one, was considered in references 
 \cite{sampaio,calegari2016}
 in a study of strongly correlated superconductors. In those works,  the good agreement between the superconducting quasiparticle bands obtained with the present methodology  and those obtained in references \cite{beenen,Haurie}, is highlighted. In addition, the superconducting order parameter as a function of electron density  also presents good agreement with that reported in references \cite{beenen,Haurie}. Therefore, we assume that the present methodology  is also suitable for dealing with the problem involving the pseudogap and the CDW in the presence of strong correlations.
 
\par This paper is organized as follows. The model and the equations obtained with the chosen method are presented in section \ref{Model}. The numerical results are presented and discussed in section \ref{results}. Conclusions and further remarks are shown in section \ref{conclusions}.

\section{The Model}\label{Model}

In order to investigate the CDW phase, we start with the following Hamiltonian \cite{balseiro}:

\begin{equation}
     \mathcal{H} =H_e+H_{PAR},
      \label{H1}
\end{equation}
with
\begin{equation}
     H_e = \sum_{\vec{k},\sigma} \xi_{\vec{k}} \ c_{\vec{k},\sigma}^{\dagger} c_{\vec{k},\sigma} 
     \label{He}
\end{equation}
and
\begin{equation}
     H_{PAR} = \sum_{\vec{k},\vec{k'},\sigma,\sigma'} V_{\vec{k}\vec{k'}} c_{{\vec{k}+\vec{Q}},\sigma}^{\dagger}c_{\vec{k},\sigma}c_{{\vec{k'}+\vec{Q}},\sigma'}^{\dagger}c_{\vec{k'},\sigma'},
      \label{Hpar}
\end{equation}
where $c_{\vec{k},\sigma}^{\dagger}(c_{\vec{k},\sigma})$ is the fermion 
creation(destruction) operator, $\sigma$ is the spin index, $V_{\vec{k}\vec{k'}}$ is the attractive pairing potential and $\xi_{\vec{k}}=\varepsilon_{\vec{k}}-\mu$, in which $\mu$ is the chemical potential. The uncorrelated dispersion relation for a square lattice is
\begin{equation}
\varepsilon_{\vek} = 2t_0[\cos{(k_xa)} + \cos{{(k_ya)}] + 4t_1 \cos{(k_xa)}\cos{(k_ya)}}, 
\end{equation}
where $a$ is the lattice constant, $t_0$ and $t_1$ are the first- and second-neighbor couplings, respectively. The commensurable wave vector $\vec{Q}$ satisfies the condition $\vec{k}+2\vec{Q}=\vec{k}$.

Following the procedure proposed by Balseiro and Falikov in reference \cite{balseiro}, the CDW is taken into account in the BCS sense. In this case, the pairing term given in Eq. (\ref{Hpar}), treated in the mean field level becomes
\begin{equation}
H_{PAR}= \sum_{\vec{k},\sigma} W_{\vec{k}} \  c_{{\vec{k}+\vec{Q}},\sigma}^{\dagger}c_{\vec{k},\sigma} + H_0,
\end{equation}
where $W_{\vek}$ is the CDW order parameter and $H_0 = \sum_{\vec{k}}|W_{\Vec{k}}|^{2}/|V| $. The 
amplitude of the order parameter is:
%
\begin{equation}
   W_{0}  =|V| \sum_{\vec{k'}} \langle c_{\vec{k'}+\vec{Q},\uparrow}^{\dagger}c_{\vek',\uparrow} \rangle, 
   \label{W0}
\end{equation}
which leads to the relation: 
\begin{equation}
    W_{\vek} = iW_{0}\gamma_{\vek},
\end{equation}
with $\gamma_{\vek} = \cos(k_xa) - \cos(k_ya)$, as the $d_{x^2 - y^2}$-wave symmetry factor. 
The parameter $V$ is the  $\vek$-independent attractive pairing potential. We propose that $W_{\vek}$ can model the CDW gap, as it represents the creation of an electron-hole pair, modulated by the vector $\vec{Q} = (\pi,\pi)$. We do not consider the Debye cutoff energy and a self-consistent non $\vek$-dependent renormalizing term \cite{balseiro}, since, in the present work,  the pairing mechanism is not considered as phononic.

The correlation function present in Eq. (\ref{W0}), can be 
obtained through the use of Green's functions in the Zubarev formalism \cite{zubarev}, whose equation of motion is: 
\begin{equation}\label{eq:mov}
    \omega \fg{\hat{A}}{\hat{B}}_{\omega} = \langle  [\hat{A},\hat{B}]_{+} \rangle + \fg{[\hat{A},\mathcal{H}]}{\hat{B}}_{\omega}.
\end{equation}

In order to obtain the set of Green's functions necessary to calculate the CDW gap amplitude $W_0$ and the chemical potential $\mu$,  the following set of operators
$\{c_{\Vec{k},\uparrow},c_{\Vec{k},\downarrow}, c_{\Vec{k}+\Vec{Q},\uparrow},c_{\Vec{k}+\Vec{Q},\downarrow}\}$ was considered. Considering the Hamiltonian given in Eq. (\ref{H1}) and applying the equation of motion to this set of operators, starting with $\fg{c_{\vec{k},\uparrow}}{c^{\dagger}_{\vec{k},\uparrow}}_{\omega}$, we have: 
\begin{equation}
         \omega \fg{c_{\vec{k},\uparrow}}{c_{\vec{k},\uparrow}^{\dagger}}_{\omega} = 1 + \xi_{\vec{k}} \fg{c_{\vec{k},\uparrow}}{c_{\vec{k},\uparrow}^{\dagger}}_{\omega} + W_{\kaq} \fg{c_{\vec{k}+\vec{Q},\uparrow}}{c_{\vec{k},\uparrow}^{\dagger}}_{\omega}.
         \label{Gcc}
\end{equation}
Repeating this procedure for the remaining operators, we obtain:
\begin{equation}
    (\omega \mathcal{I} - \mathcal{H}_{\vec{k}})\mathcal{G}(\vek,\omega)  = \mathcal{I},
\label{4}
\end{equation}
where $\mathcal{G}_{\vek}(\vek,\omega)$ is the Green's function matrix, whose elements are given by $\mathcal{G}_{ls}=\fg{\hat{A}_l}{\hat{B}_s}$. $\mathcal{I}$ is a fourth-order identity matrix and: 
\begin{equation}
    \mathcal{H}_{\vek} = 
\begin{pmatrix}\label{eq:matrizCoeff}
\xi_{\vek} & 0 & W_{\vek} & 0 \\ 
0 & -\xi_{\vek} & 0 & -W_{\vek}^{*} \\ 
W_{\vek}^{*} & 0 & \xi_{\veq} & 0  \\
0 & -W_{\vek} & 0 & -\xi_{\veq}
\end{pmatrix},
\end{equation}
in which it was assumed that $W_{\vek} = W_{\veq}^{*}$.

In the normal state $W_{\ka} = 0$, therefore Eq. (\ref{Gcc}) reduces to 
\begin{equation}
    G_0(\ka,\omega) = (\omega - \xi_{\vek})^{-1},
\label{G0}
\end{equation}
where $G_0(\ka,\omega) =\fg{c_{\vec{k},\uparrow}}{c^{\dagger}_{\vec{k},\uparrow}}_{\omega}$
is the uncorrelated Green's function for the normal state. In terms of $G_0(\ka,\omega) $, the quantity $\omega \mathcal{I} - \mathcal{H}_{\vec{k}}$ present in Eq. (\ref{4}), can be written as:
\begin{equation}
    \omega \mathcal{I} - \mathcal{H}_{\vec{k}}= 
\begin{pmatrix}\label{eq:matrizC}
G_{0}^{-1}(\ka) & 0 & W_{\vek} & 0 \\ 
0 & -G_{0}^{-1}(\ka)  & 0 & -W_{\vek}^{*} \\ 
W_{\vek}^{*} & 0 & G_{0}^{-1}(\ka')  & 0  \\
0 & -W_{\vek} & 0 & -G_{0}^{-1}(\ka') 
\end{pmatrix},
\end{equation}
where $\ka'=\vec{k}+\vec{Q}$ and the dependence of $G_{0}$ on $\omega$ was omitted. 

In order to introduce the correlation effects due to the Coulomb interaction, 
at this point, the Green's function $G_0(\vec{k},\omega)$  is replaced  by a normal state correlated Green's function obtained through an n-pole approximation \cite{Roth,beenen}, applied to the one-band Hubbard model \cite{Hubbard}. As in the study of superconductivity in references \cite{sampaio,calegari2016,RODRIGUEZNUNEZ200188}, we are assuming that the presence of correlations does not significantly affect the BCS formalism.

The two-dimensional one-band Hubbard model is:
\begin{equation}
    H = \sum_{\langle \langle i,j\rangle \rangle,\sigma }t_{ij}c_{i,\sigma}^{\dagger}c_{j,\sigma} + \frac{U}{2}\sum_{i,\sigma}n_{i,\sigma}n_{i,-\sigma},
\end{equation}
where $c_{i,\sigma}^{\dagger}(c_{i,\sigma})$ is the fermion creation(destruction) operator and $n_{i,\sigma} = c_{i,\sigma}^{\dagger}c_{i,\sigma}$ is the occupation number operator with spin $\sigma = \{\uparrow,\downarrow \}$. The first term in the Hamiltonian $H$ describes the hopping of the electrons through the lattice sites where the symbol $\langle \langle ... \rangle \rangle$ indicates a sum over first- and second-nearest neighbors. The second term in $H$ considers the repulsive Coulomb interaction between two electrons with opposite spins located at the same lattice site $i$. 
We highlight that within the n-pole approximation, the second-nearest neighbor hopping is crucial to properly capture the short-range antiferromagnetic correlation effects that give rise to the pseudogap phenomenon.

In the n-pole approximation \cite{Roth,beenen},  
the correlated Green's function matrix is given by:
 \begin{equation}
       \textbf{G}(\vek,\omega) = \textbf{N}\left(\omega\textbf{N}-\textbf{E}(\vec{k})\right)^{-1}\textbf{N}, 
   \end{equation}
where $N_{nm} = \langle [ \hat{A}_n,\hat{A}^{\dagger}_m]_{+} \rangle$ and $E_{nm} = \langle [[\hat{A}_n,\mathcal{H}],\hat{A}^{\dagger}_m]_{+} \rangle$ are the elements of the normalization and the energy matrices, respectively. We follow Roth \cite{Roth} in choosing the set of operators $\{\hat{A}_n\} = \{ c_{i,\sigma}, n_{i,-\sigma}c_{i,\sigma}\} $ to describe the correlated normal state.
One of the most important elements of $\textbf{G}(\vek,\omega)$ is given by: 
\begin{equation}\label{eq:roth}
G_{11,\sigma}(\omega,\vek) = \frac{\omega - U(1-\langle n_{-\sigma}\rangle) - Y_{\vec{k},-\sigma}}{(\omega - \varepsilon_{\vek})(\omega - U - Y_{\vec{k},-\sigma}) - U \langle n_{-\sigma}\rangle(\varepsilon_{\vec{k}} - Y_{\vec{k},-\sigma})},
\end{equation}
in which $\langle n_\sigma\rangle=\langle n_{i,\sigma}\rangle$ is the average site occupation per spin and $Y_{\vec{k},\sigma}$ is the band shift defined as:
\begin{equation}
    \langle n_{\sigma}\rangle (1 - \langle n_{\sigma}\rangle) Y_{\vec{k},\sigma} = \Gamma^{(0)}_{\sigma} + \sum_{j \neq i} e^{i\vec{k}\cdot (\vec{R}_i - \vec{R}_j)}t_{ij} \big( \Gamma^{(1)}_{i,j,\sigma} + \Gamma^{(2)}_{i,j,\sigma} + \Gamma^{(3)}_{i,j,\sigma} \big).
\end{equation}
The quantities $\Gamma^{(m)}_{i,j,\sigma}$ are given by:
\begin{equation}
    \Gamma^{(0)}_{\sigma} = -\sum_{j \neq i} t_{ij} \langle c_{i,\sigma}^{\dagger}c_{j,\sigma}(1 -  n_{i,-\sigma} - n_{j, -\sigma}) \rangle,
    \label{gama0}
\end{equation}
\begin{equation}
    \Gamma^{(1)}_{i,j,\sigma} = \frac{1}{4} (\langle N_i N_j \rangle - \langle N_j \rangle \langle N_i \rangle), 
    \label{gama1}
\end{equation}

\begin{equation}
     \Gamma^{(2)}_{i,j,\sigma} =  \langle \vec{S}_{i} \cdot \vec{S}_{j} \rangle
    \label{gama2}
\end{equation}
\noindent and 
\begin{equation}
     \Gamma^{(3)}_{i,j,\sigma} = - \langle c_{j,\sigma}^{\dagger}c_{j,-\sigma}^{\dagger} c_{i,-\sigma} c_{i,\sigma} \rangle,
    \label{gama3}
\end{equation}
where $N_i = n_{i,\sigma} + n_{i,-\sigma}$ is the total number operator per site while $\vec{S}_i$ is the spin operator. In order to properly take into account the short-range antiferromagnetic correlations which are responsible for the pseudogap,
it is  essential to adequately include the momentum dependence of $Y_{\vec{k},\sigma}$ \cite{calegari2011}.
Thus, it
is rewritten as:
\begin{equation}
        \langle n_{\sigma} \rangle (1 - \langle n_{\sigma} \rangle) Y_{\vec{k},\sigma} = \Gamma^{(0)}_{\sigma} + \sum_{\vec{q}}\sum_{m = 1}^{3}  \varepsilon(\Vec{k} - \Vec{q}) \Gamma^{(m)}_{\vec{q},\sigma},
\end{equation}
where $\Gamma^{(m)}_{\vec{q},\sigma}$ is the inverse Fourier transform of $\Gamma^{(m)}_{i,j,\sigma}$,
        \begin{equation}
            \Gamma^{(m)}_{\vec{q},\sigma} = \frac{1}{L} \sum_{\Vec{i,j}} e^{i\vec{q}\cdot(\Vec{R}_j - \Vec{R}_i)} \Gamma_{i,j,\sigma}^{(m)}
            \end{equation}
and the dispersion relation is 
\begin{equation}
            \varepsilon(\Vec{k} - \Vec{q}) = \frac{1}{L} \sum_{j \neq i} e^{i(\Vec{k} - \Vec{q})\cdot(\Vec{R}_j - \Vec{R}_i)}t_{ij},
\end{equation}
in which $L$ is the number of lattice sites.
The correlation functions given in Eqs. (\ref{gama0})-(\ref{gama3}), have been calculated following the original procedure proposed by Roth \cite{Roth}.

As discussed before, in order to take into account correlations in the CDW state, the uncorrelated Green's function $G_{0}(\vec{k},\omega)$ present in Eq. (\ref{eq:matrizC}), was replaced by the correlated Green's function $G_{11,\sigma}(\vec{k},\omega)$, given in Eq. (\ref{eq:roth}). Using the resulting $\omega \mathcal{I} - \mathcal{H}_{\vec{k}}$ to solve Eq. (\ref{4}), we obtain the correlated Green's function matrix $\mathcal{G}(\vek,\omega)$ for the CDW state. 
Writing the elements of $\mathcal{G}(\vek,\omega)$ in terms of partial fractions,
the general form for these Green's functions is 
\begin{equation}\label{eq:pesos}
    \mathcal{G}_{ls,\sigma}(\vec{k},\omega) = \sum_{m=1}^{4}\frac{Z^{(ls)}_{m,\sigma}(\vec{k})}{\omega - E_{m}(\vec{k})},
\end{equation}
where $Z^{(ls)}_{m,\sigma}$ and  $E_{m}(\vec{k})$ are the spectral weights and the quasiparticle  bands, respectively.
\begin{figure}[!t]
\begin{center}
\leavevmode
\includegraphics[width=0.65\linewidth]{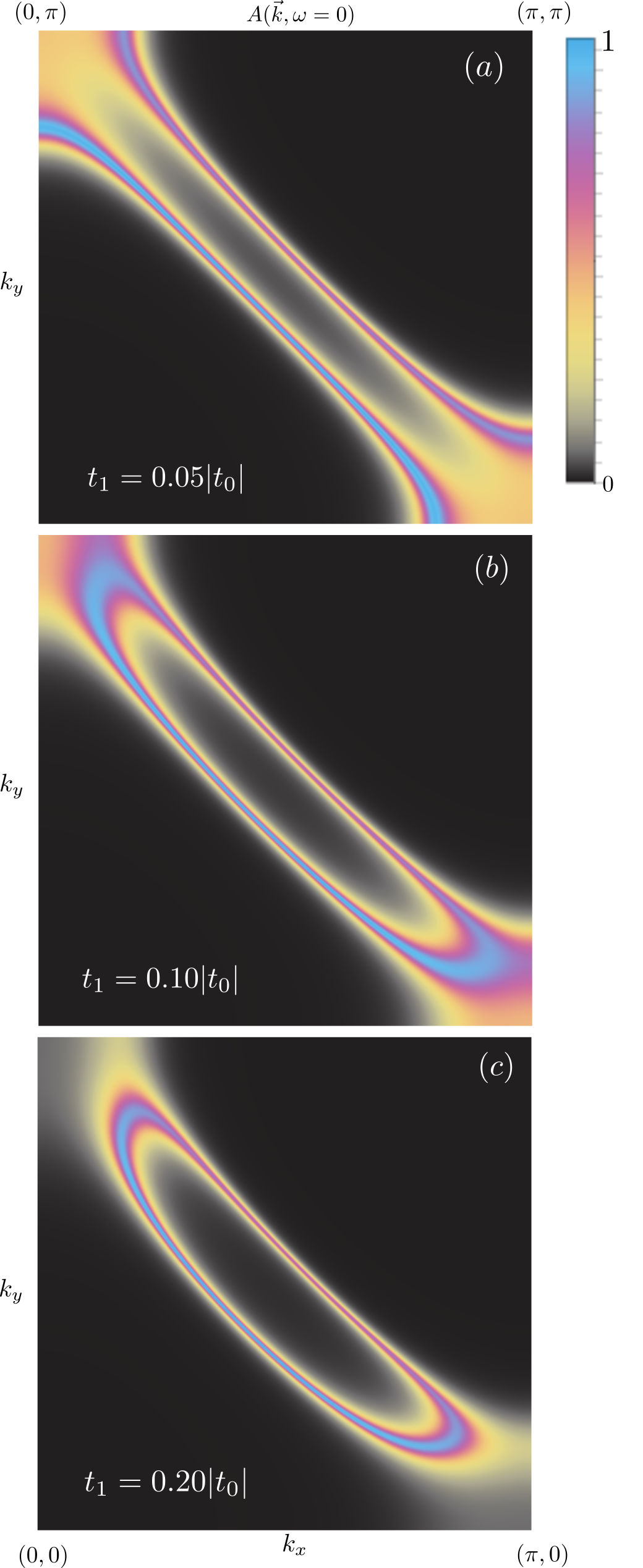}
\end{center}
\caption{Spectral function $A_{\sigma}(\vec{k},\omega=0)$ showing the evolution of the Fermi surface in the normal state with $U = 9.0|t_0|$, $n_T = 0.90$ and $k_BT = 0.05|t_0|$, for (a) $t_1 = 0.05|t_0|$, (b) $t_1 = 0.10|t_0|$ and (c) $t_1 = 0.20|t_0|$. In Fig. (c), the Fermi surface collapses into a pocket, indicating the opening of a pseudogap.}
\label{fig:r10}
\end{figure}
One of the most relevant Green's functions for this work is 
$\mathcal{G}_{11,\sigma}(\vek,\omega)$, which is associated with the occupation number $\langle n_{-\sigma}\rangle$, the DOS and the spectral function. However, $ \mathcal{G}_{13,\sigma}(\vek,\omega)$ enables the calculation of the CDW order parameter. To evaluate the necessary correlation functions, we use the standard relation  
\begin{equation}
    \langle \hat{B} \hat{A} \rangle = \oint \langle \langle \hat{A} ; \hat{B} \rangle \rangle_{\omega} \eta_F(\omega) d\omega,
     \label{cf}
\end{equation}
where the contour encircles the real axis without enclosing the poles of the Fermi
function $\eta_F(\omega)$. Applying that representation to $ \mathcal{G}_{11,\sigma}(\vek,\omega)=\fg{c_{\vec{k},\sigma}}{c_{\vec{k},\sigma}^{\dagger}}_{\omega}$ results in the average of the occupation number per spin:

\begin{equation}
    \langle n_{-\sigma} \rangle = \frac{1}{L}\sum_{\vec{k}}\sum_{m=1}^{4} Z^{(11)}_{m,\sigma}(\vec{k})\eta_{F}\left(E_{m}(\vec{k})\right).
    \label{ns}
\end{equation}
In terms of $\mathcal{G}_{11,\sigma}(\vek,\omega)$, the  spectral function $A_{\sigma}(\vec{k},\omega)$ is:
\begin{equation}
A_{\sigma}(\vec{k},\omega)=-\frac{1}{\pi}\operatorname{Im} [\mathcal{G}_{11,\sigma}(\vek,\omega)].
\label{Ak}
\end{equation}

\par Considering  $ \mathcal{G}_{13,\sigma}(\vek,\omega)$ and the relation introduced in Eq. (\ref{cf}), the amplitude of the CDW order parameter given in Eq. (\ref{W0}) is  rewritten as: 
\begin{equation}
     W_{0} =\frac{|V|}{L} \sum_{\vec{k}} \sum_{m=1}^{4} Z^{(13)}_{m,\sigma}(\vec{k})\eta_{F}\left(E_{m}(\vec{k})\right).
     \label{W01}
\end{equation}

\section{Numerical Results}\label{results}

To ensure the consistency of the results, we use $|t_0|$ as the energy unit, with  $t_0 = -1.0$ eV. The attractive pairing potential  $V = 1.2t_0$ was kept fixed for all results presented in this section. Furthermore, $n_T = \langle n_{\sigma}\rangle + \langle n_{-\sigma}\rangle$ is the total occupation per site.
In this work, the equations (\ref{ns}) and (\ref{W01}) have been self-consistently solved in order to obtain, respectively,
the chemical potential $\mu$ and the CDW gap amplitude $W_0$ for a given temperature and a set of model parameters. It has been shown in references \cite{beenen,calegari2005} that the correlation functions present in the band shift 
 $Y_{\vek,\sigma}$ are not significantly affected by the ordered phase. Therefore, in the present work the correlation functions composing $Y_{\vek,\sigma}$ were evaluated only in the normal state.

\par The  spectral function $A_{\sigma}(\vec{k},\omega=0)$ in the normal state is shown in Fig. \ref{fig:r10} for different values of $t_1$.
The result allows an analysis of the evolution of the Fermi surface in terms of $t_1$. In Fig. \ref{fig:r10}(a), the region in which $A_{\sigma}(\vec{k},\omega=0)$ is more intense defines the main electron-like Fermi surface centered at $(0,0)$. Figures \ref{fig:r10}(b) and \ref{fig:r10}(c), show the Fermi surface evolving into a  pocket, when $t_1$ increases. In Fig. \ref{fig:r10}(c), the low spectral intensity near the nodal points  $(0,\pi)$ and  $(\pi,0)$ indicates the presence of pseudogaps at these regions \cite{Avella}. It is important to note that before the opening of the pseudogap, the Fermi surface undergoes a Lifshitz transition for some value of $t_1$ between $0.05|t_0|$ and $0.10|t_0|$. Therefore, the opening of the pseudogap induced by 
$t_1$, is preceded by a Lifshitz transition \cite{Chen2012}.

\begin{figure}[!ht]
    \begin{center}
    \leavevmode
    \includegraphics[width=0.85\linewidth]{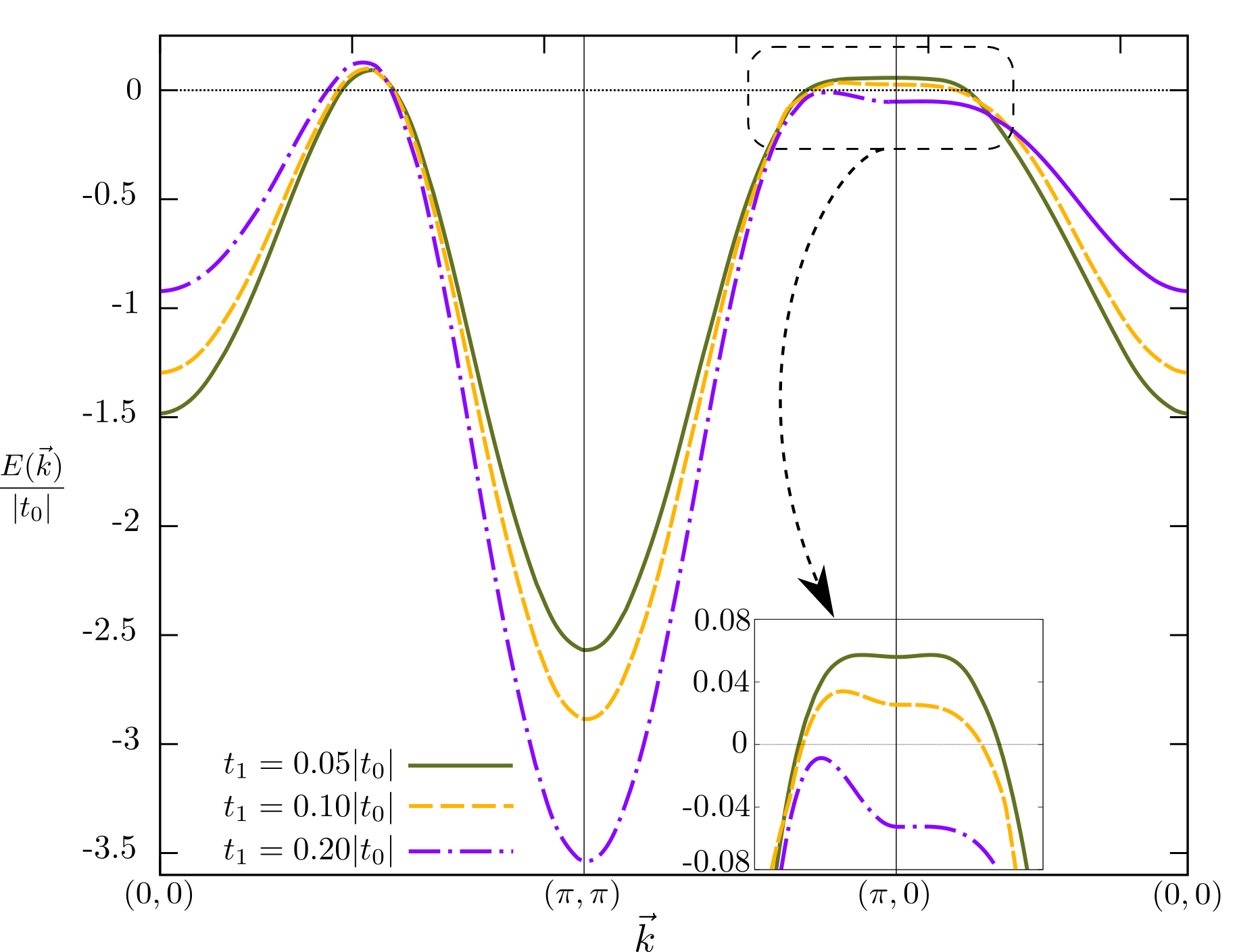}
    \end{center}
    \caption{The normal state quasiparticle bands ($E_1(\vec{k})$ ) intercepted by the chemical potential for $U = 9.0|t_0|$, $n_T = 0.90$, $k_BT = 0.05|t_0|$ and different values of $t_1$.
    The inset shows in details the antinodal point region where the pseudogap emerges. The dotted black line indicates the position of the chemical potential.}
    \label{fig:r11}
\end{figure}

In the present scenario, the pseudogap arises due to short-range antiferromagnetic correlations, which are enhanced by the increasing of $t_1$.
The mechanism is characterized by a distortion of the quasiparticle bands because of the effect of the short-range antiferromagnetic correlations which 
displace the quasiparticle band at the region of the nodal point $(\pi,\pi)$  to lower energies, as can be seen in Fig. \ref{fig:r11}. However, a portion of the quasiparticle band around the point $(\pi/2,\pi/2)$ remains above the chemical potential level, while at the region of the antinodal point  $(\pi,0)$, the quasiparticle band lies  below the chemical potential, when $t_1$ reaches a certain value.
This behavior gives rise to a pseudogap at $(\pi,0)$, as can be seen in the inset of Fig. \ref{fig:r11}, for $t_1=0.20|t_0|$.

\begin{figure}[!ht]
    \begin{center}
    \leavevmode
    \includegraphics[width=0.85\linewidth]{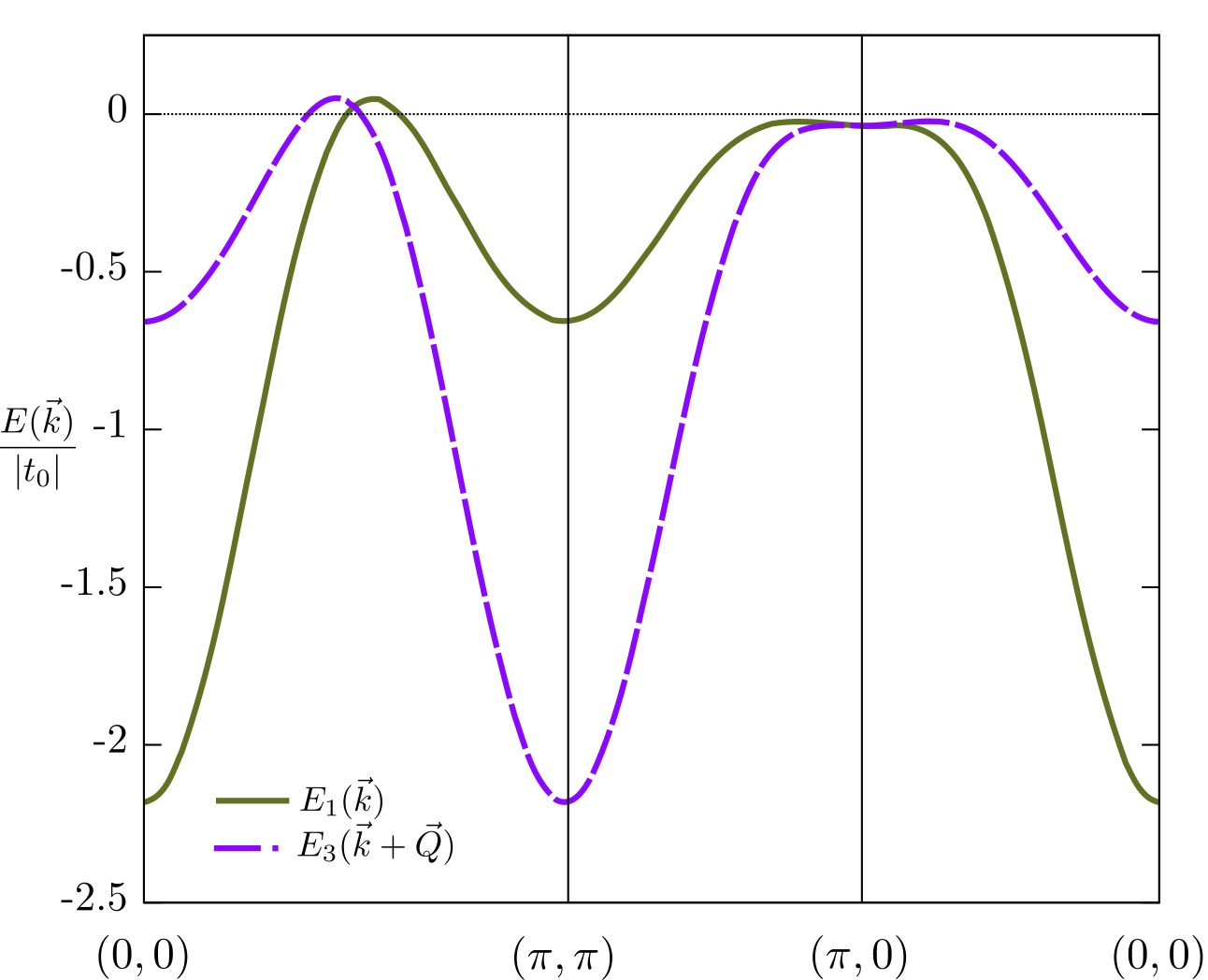}
    \end{center}
    \caption{The quasiparticle bands $E_1(\vec{k})$ and $E_3(\vec{k}+\vec{Q})$ intercepted by the chemical potential for $U = 16.0|t_0|$, $n_T = 0.90$, $t_1 = 0.12|t_0|$
    and $k_BT = 0.15|t_0|$. The dotted black line indicates the position of the chemical potential.}
    \label{fig:r101}
\end{figure}
\begin{figure}[!ht]
\begin{center}
\leavevmode
\includegraphics[width=0.75\linewidth]{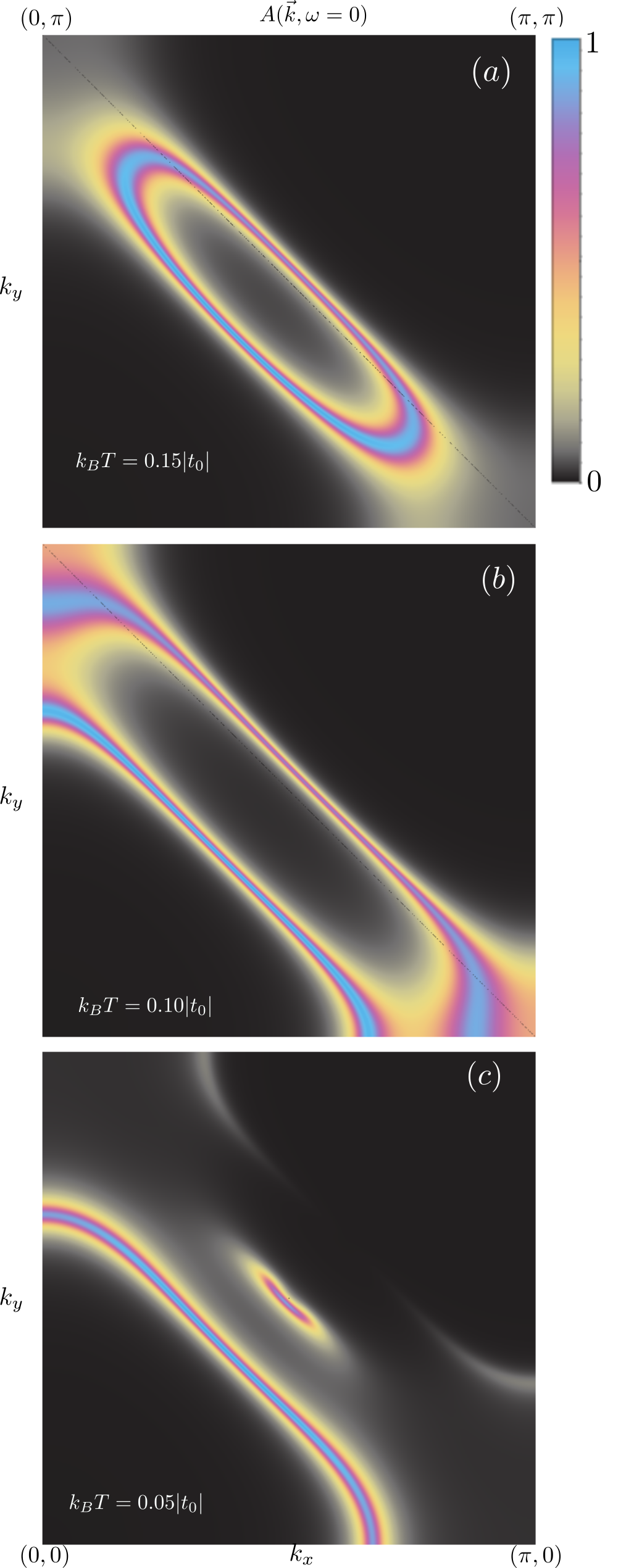}
\end{center}
\caption{The spectral function $A_{\sigma}(\vec{k},\omega=0)$ for different values of $k_BT$.
(a) For $k_BT = 0.15|t_0|$, we observe the presence of a pocket in the normal state. In (b),  the pocket falls apart converting into two electron-like Fermi surfaces. In (c), for $k_BT = 0.05|t_0|$, the system lies in the CDW phase and only a portion of the fermi surface remains.  The considered model parameters are  $U = 16|t_0|$, $n_T = 0.90$ and $t_1 = 0.12|t_0|$.}
 \label{fig:r6}
\end{figure}
\begin{figure}[!t]
    \begin{center}
    \leavevmode
    \includegraphics[width=\linewidth]{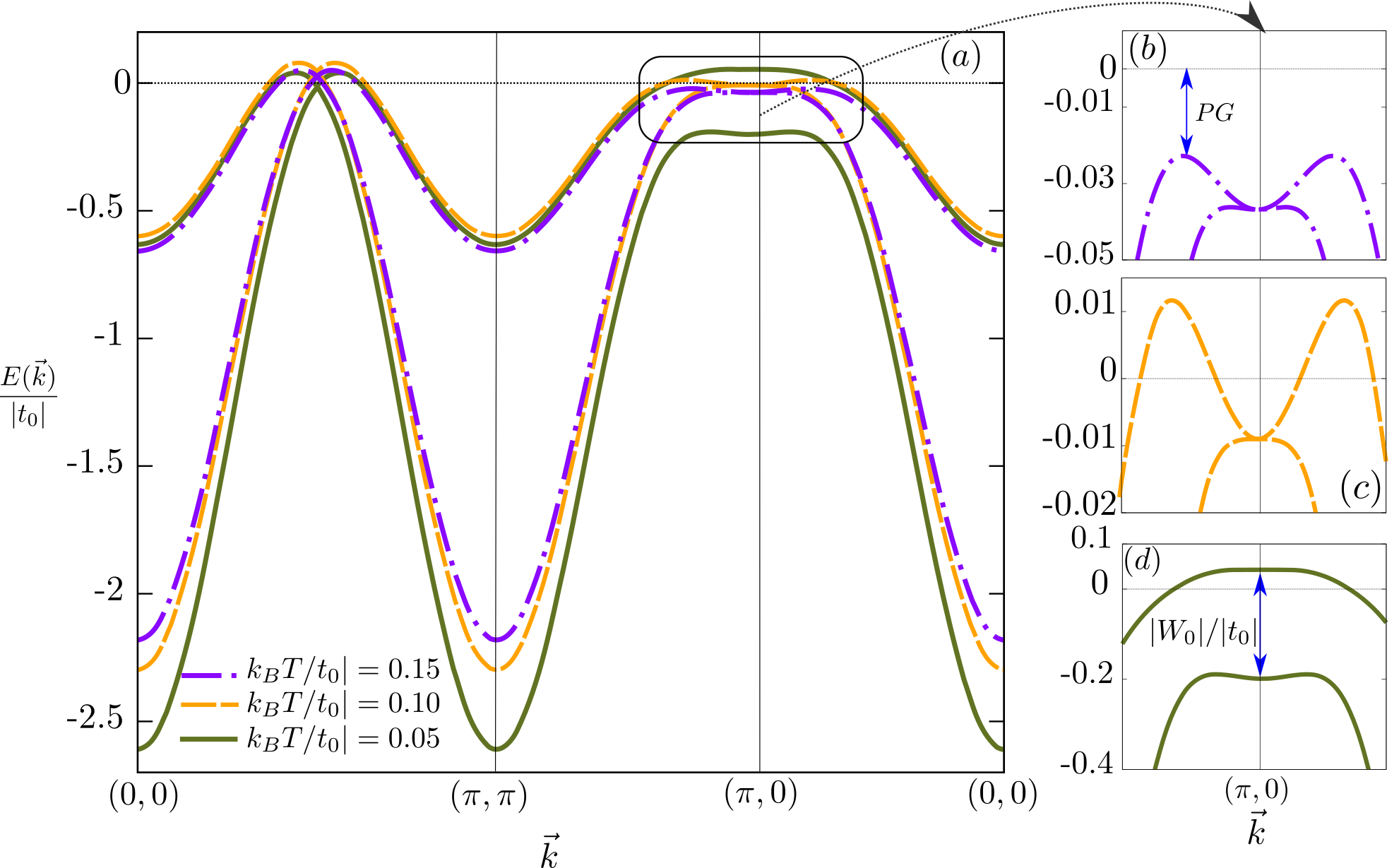}
    \end{center}
    \caption{(a)  Quasiparticle bands for $U = 16.0|t_0|$, $n_T = 0.90$, $t_1 = 0.12|t_0|$ and different values of $k_BT$. (b) The quasiparticle bands close to the chemical potential level showing in detail the pseudogap at the antinodal point $(\pi,0)$, for $k_BT = 0.15|t_0|$. (c) Close-up view near the antinodal point for $k_BT = 0.10|t_0|$. In (d), the CDW gap for $k_BT = 0.05|t_0|$.  The dotted black lines indicate the position of the chemical potential.}
    \label{fig:r7}
\end{figure}

To investigate the CDW phase, it is necessary to also consider the quasiparticle bands which depend on $\vec{k}+\vec{Q}$, where $\vec{Q}=(\pi,\pi)$ is the CDW modulation vector.
The effect of the short-range antiferromagnetic correlations on the  quasiparticle bands $E_1(\vec{k})$ and $E_3(\vec{k}+\vec{Q})$ can be verified in Fig. \ref{fig:r101}. It is important to remember that in the present methodology, the short-range antiferromagnetic correlations are associated with the spin-spin correlation function $\langle \vec{S}_{i} \cdot \vec{S}_{j} \rangle$, present in the band shift $Y_{\vec{k},\sigma}$. As discussed for the result depicted in Fig. \ref{fig:r11}, the short-range antiferromagnetic correlations shift the quasiparticle band $E_1(\vec{k})$ to low energies in the vicinity of the nodal point $(\pi,\pi)$, enabling the emergence of a pseudogap at the antinodal point $(\pi,0)$. Nevertheless, in the case of $E_3(\vec{k}+\vec{Q})$, it is the region near to the
origin $(0,0)$ which is shifted to low energies.
Indeed, the effect of  the short-range antiferromagnetic correlations on $E_1(\vec{k})$ and $E_3(\vec{k}+\vec{Q})$ is responsible for a rich Fermi surface topology, mainly in the CDW phase.

\par Figure \ref{fig:r6} shows the evolution of the Fermi surface with $k_BT$. In Fig. \ref{fig:r6}(a), $k_BT = 0.15|t_0|$ and the system is found in the normal state. In this case, we observe a pocket centered at the point $(\pi/2,\pi/2)$, agreeing with the result shown in Fig. \ref{fig:r10}(c) for the normal state. If the temperature decreases to $0.10|t_0|$, the pocket evolves to two electron-like Fermi surfaces as shown in Fig. \ref{fig:r6}(b). When the temperature drops to $0.05|t_0|$, the system enters the CDW phase
and an electron-like Fermi surface is still present, differing from the expected behavior of a total Fermi surface collapse. To better understand the Fermi surface behavior, we analyzed the evolution of the quasiparticle bands in terms of the temperature, as shown in  Fig. \ref{fig:r7}. Figs. \ref{fig:r7}(b)-\ref{fig:r7}(d), show details of the antinodal point region for different temperatures. The result for the normal state with $k_BT=0.15|t_0|$ is shown in Fig. \ref{fig:r7}(b), in which a pseudogap is observed, as the bands do not cross $\mu$ at $(\pi,0)$, but intersections still occur at $(\pi/2, \pi/2)$, as seen in \ref{fig:r7}(a). The result depicted in Fig. \ref{fig:r7}(c), shows that, due to the effect of short-range antiferromagnetic correlations, the quasiparticle bands $E_1(\vec{k})$ and $E_3(\vec{k}+\vec{Q})$   
crosses  the chemical potential $\mu$ twice in the regions of both points $(\pi/2,\pi/2)$ and $(\pi,0)$.
As a consequence, two concentric electron-like Fermi surfaces are observed in Fig. \ref{fig:r6}(b). For  $k_BT=0.05|t_0|$, the system lies in the CDW phase with a 
well-defined gap at the antinodal point $(\pi,0)$,  as can be seen in Fig. \ref{fig:r7}(d). However, because of the effect of the short-range antiferromagnetic correlations, the upper band in Fig. \ref{fig:r7}(d) crosses the chemical potential even in the CDW state, resulting in a reconstruction of an electron-like Fermi surface as shown in Fig.  \ref{fig:r6}(c).

\begin{figure}[!th]
    \begin{center}
    \leavevmode
    \includegraphics[width=0.8\linewidth]{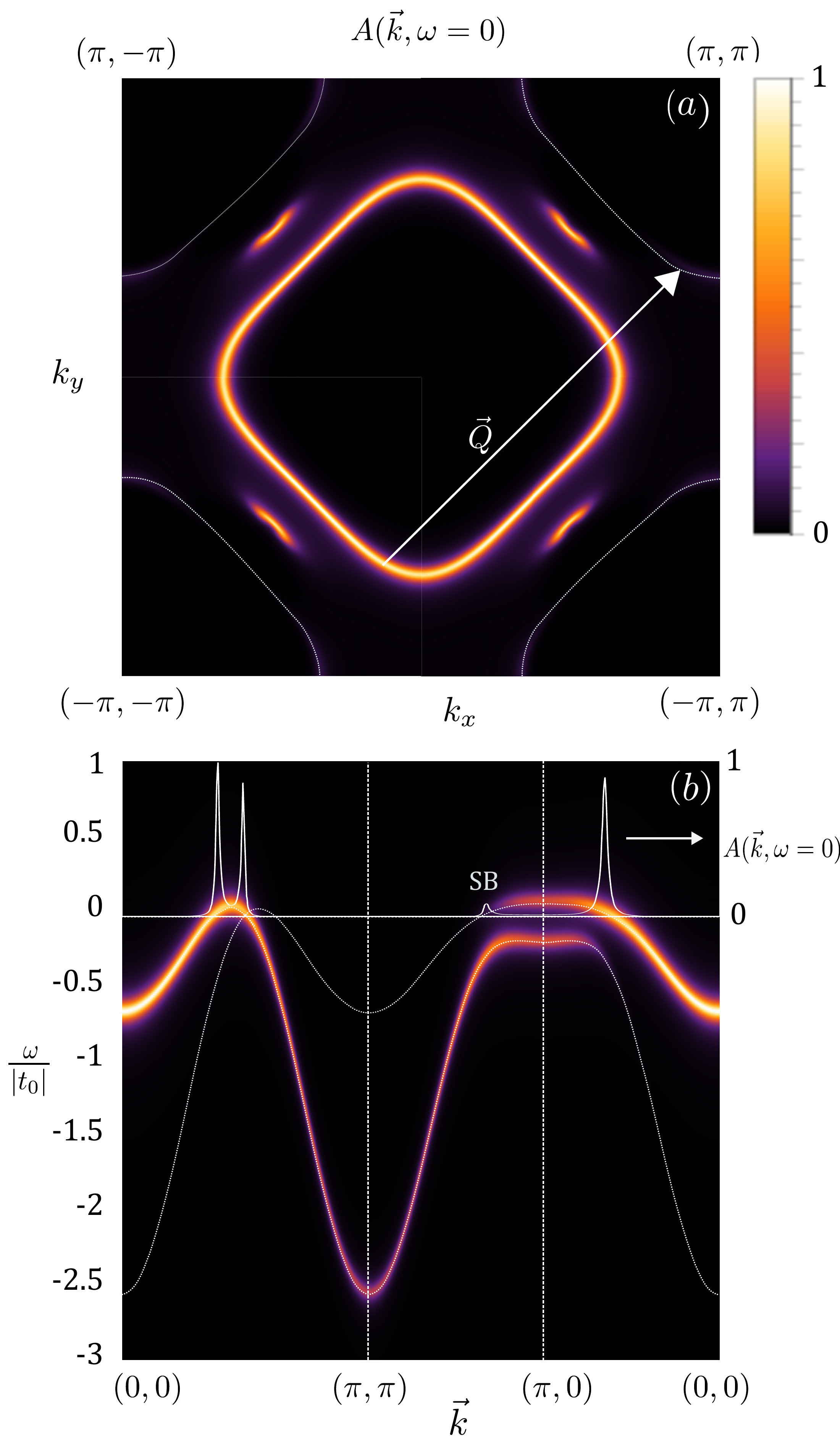}
    \end{center}
    \caption{(a) The spectral function $A(\vec{k},\omega=0)$ in the CDW phase with  $U = 16.0|t_0|$, $n_T = 0.90$, $t_1 = 0.1|t_0|$ and $k_BT=0.05|t_0|$. (b) The spectral function $A(\vec{k},\omega)$ along the main directions is shown for the same model parameters as in (a). The white solid line shows the spectral function $A(\vec{k},\omega=0)$, while the  horizontal dotted white line at $\omega=0$ indicates the position of the chemical potential.}
    \label{fig:r3}
\end{figure}

The Fermi surface topologies are closely-related to important physical properties which can help to better understand exotic phenomena like the pseudogap. Fig. \ref{fig:r3}(a)  displays the Fermi surface  for the CDW phase with $n_T = 0.90$.
Note the existence of a main electron-like Fermi surface centered at $(0,0)$, which emerges due to the crossing of the upper band and the chemical potential, as shown in Fig. \ref{fig:r7}(d). However, it is also possible to see the remains of a hole-like Fermi surface centered at $(\pi,\pi)$ which is characterized by a low spectral intensity. This last Fermi surface may be associated with a shadow band (SB) which arises due to antiferromagnetic correlations  that are more intense in the strongly correlated regime. In Fig. \ref{fig:r3}(a), the vector $\vec{Q}=(\pi,\pi)$   shows the perfect nesting between both the main and the SB Fermi surfaces.
The relation between the shadow bands and the antiferromagnetic correlations was the object of theoretical studies \cite{kampf,moreo,vilk1997shadow}. Moreover, there are also experimental data reporting shadow bands in cuprates \cite{nakayama,Mannella} and recently, in the CDW material CuTe \cite{Zhong}. In Fig. \ref{fig:r3}(a),  it is also possible to see a small "arc" at $(\pi/2,\pi/2)$ reminiscent of the pocket present in the normal state.
Figure \ref{fig:r3}(b) shows the spectral function $A(\vec{k},\omega)$ along the main directions for the same model parameters considered in Fig. \ref{fig:r3}(a). The width of the curves and the color map indicate the intensity of the spectral function in terms of vector $\vec{k}$. The white solid line shows the spectral intensity $A(\vec{k},\omega=0)$, in which it is possible to identify a small peak associated to the shadow band responsible for the hole-like Fermi surface shown in \ref{fig:r3}(a).  

\begin{figure}[!th]
    \begin{center}
    \leavevmode
    \includegraphics[width=0.8\linewidth]{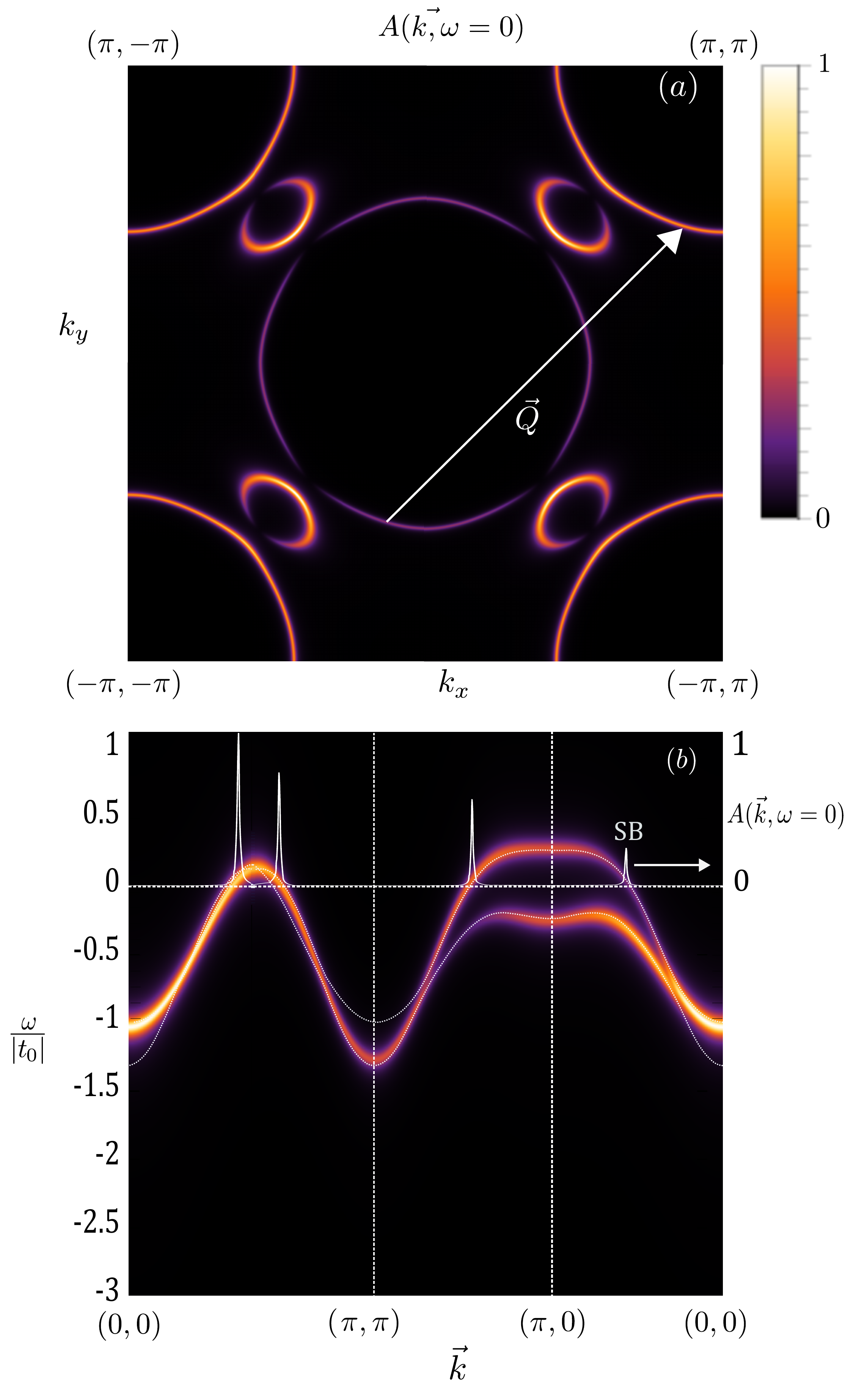}
    \end{center}
    \caption{(a) The spectral function $A(\vec{k},\omega=0)$ in the CDW phase for $n_T = 0.85$,  $U = 16.0|t_0|$, $t_1 = 0.1|t_0|$ and $k_BT=0.05|t_0|$. (b) The spectral function $A(\vec{k},\omega)$ along the main directions is shown for the same model parameters as in (a). The white solid line shows the spectral function $A(\vec{k},\omega=0)$, while the  horizontal dotted white line at $\omega=0$ indicates the position of the chemical potential.}
    \label{fig:r4}
\end{figure}

It is interesting to note that the Fermi surface topology changes significantly when the occupation decreases to $n_T=0.85$. As shown in Fig. \ref{fig:r4}(a), the hole-like Fermi surface centered at $(\pi,\pi)$ now presents a higher spectral intensity when compared to the electron-like Fermi surface centered at $(0,0)$.  This behavior indicates a transfer of spectral weight from the region of the direction $(\pi,0)$-$(0,0)$ to the region of the direction $(\pi,0)$-$(\pi,\pi)$, as can be seen in Fig. \ref{fig:r4}(b). This occurs due to the weakening of the short-range antiferromagnetic correlations caused by the decreasing of the electron density.
The vector $\vec{Q}=(\pi,\pi)$  shows the perfect nesting between both the main and the SB Fermi surfaces. Besides, the  "arc" at $(\pi/2,\pi/2)$ reminiscent of the pocket present in the normal state, is more evident now.
Fig. \ref{fig:r4}(b) shows the same result as in  \ref{fig:r3}(b), but now for  $n_T=0.85$. Due to the weakening of the short-range antiferromagnetic correlations caused by the decreasing of the electron density, the quasiparticle bands $E_1(\vec{k})$ and $E_3(\vec{k}+\vec{Q})$ are less shifted towards low energies in the regions of the points $(\pi,\pi)$ and $(0,0)$, when compared with the result shown in \ref{fig:r3}(b). If we compare the results for the Fermi surface shown in Figs. \ref{fig:r3}(a) and \ref{fig:r4}(a), we notice that there is a Lifshitz transition in the main Fermi surface. For $n_T=0.90$ there is an electron-like Fermi surface which evolves to a hole-like Fermi surface when  $n_T=0.85$.

\begin{figure}[!h]
    \begin{center}
    \leavevmode
    \includegraphics[width=0.75\linewidth]{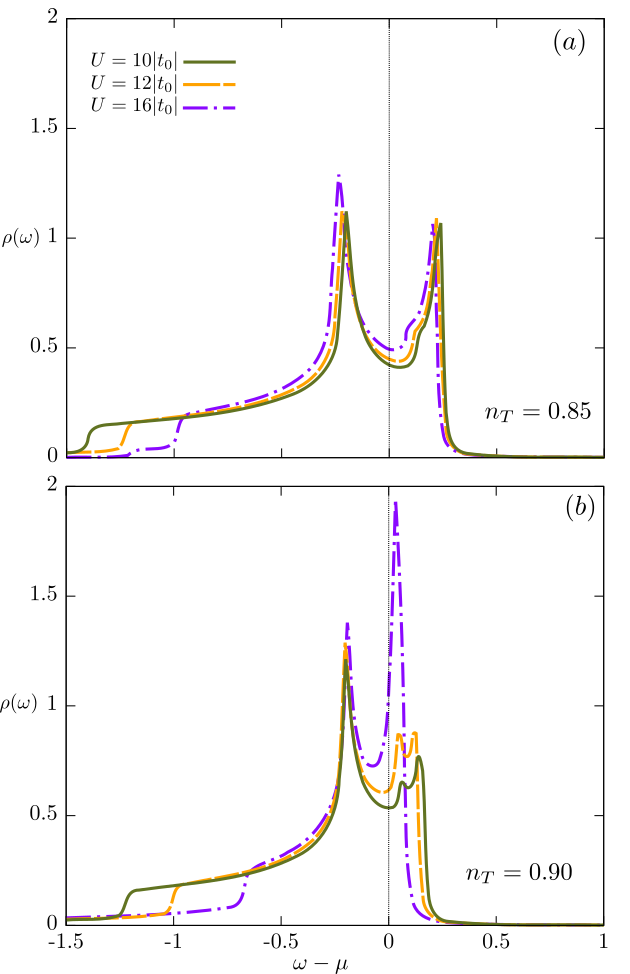}
    \end{center}
    \caption{The density of states in the CDW phase for (a) $n_T = 0.85$ and (b) $n_T = 0.90$. The next nearest neighbor hopping and temperature are $t_1=0.1|t_0|$ and $k_BT=0.05|t_0|$, respectively. The dotted black line indicates the position of the chemical potential.}
    \label{fig:r5}
\end{figure}

Since in the present case, 
the CDW phase exhibits a rich Fermi surface topology, it is also worth analyzing the behavior of the DOS which is shown in Fig. \ref{fig:r5} for  $n_T = 0.85$ and $n_T = 0.90$.
For a CDW phase with $d_{x²-y²}$-wave symmetry, we would expect a DOS with a full gap at $\omega=0$, however, a partial gap is observed. This occurs due to the contribution of the states coming from the upper quasiparticle band which cross the chemical potential twice in the directions $(\pi,0)$-$(0,0)$ and $(\pi,0)$-$(\pi,\pi)$, as can be seen in Fig. \ref{fig:r4}(b). As discussed earlier, this band behavior is directly related to the short-range antiferromagnetic correlations, which move the quasiparticle bands $E_1(\vec{k})$ and $E_3(\vec{k}+\vec{Q})$ to lower energies at the points $(\pi,\pi)$ and $(0,0)$, respectively. 
For $n_T=0.90$, the effect of the correlations becomes stronger, the depth of the partial gap decreases and the system approaches a pseudogap regime in the normal state.

\begin{figure}[!h]
    \begin{center}
    \leavevmode
    \includegraphics[width=0.8\linewidth]{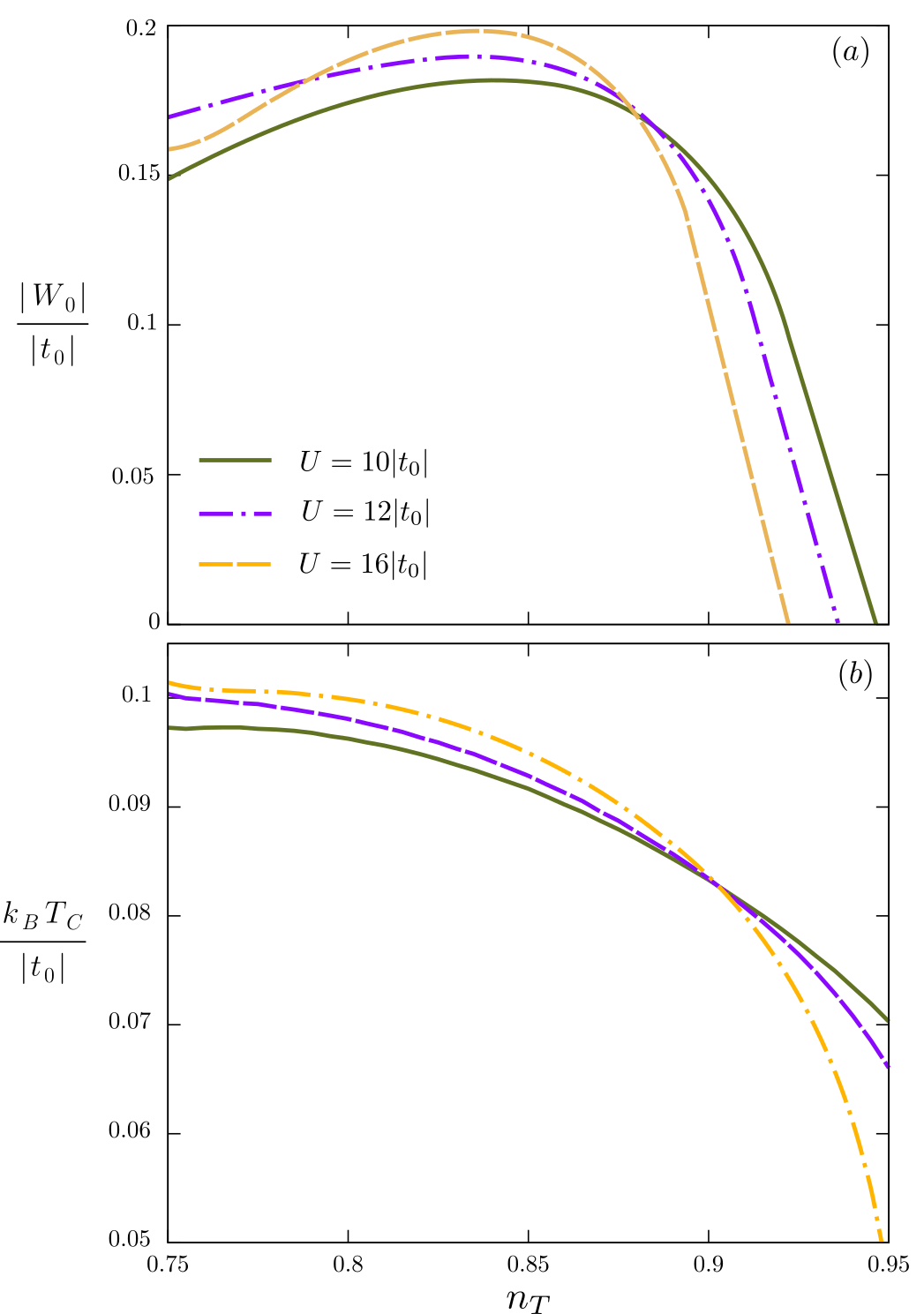}
    \end{center}
    \caption{(a) Amplitude of the CDW order parameter and (b) critical temperature as a function of $n_T$ for $t_1 = 0.1|t_0|$, $k_BT = 0.05|t_0|$ and different values of $U$.}
    \label{fig:r13}
\end{figure}

The amplitude of the order parameter as a function of $n_T$ is shown in Fig. \ref{fig:r13}(a) for different values of $U$. An initial favoring is followed by suppression in the order parameter amplitude, caused by increasing $n_T$. This implies an optimal correlation regime for $n_T\sim 0.84$, where the antiferromagnetic correlations favor CDW stability. However, if $n_T$ keeps increasing, the system enters a regime of stronger correlations in which the antiferromagnetic correlations are enhanced and the CDW is now suppressed.
To further investigate the CDW regime,  $k_BT_C$ as a function of $n_T$ is shown in Fig. \ref{fig:r13}(b),
for various values of $U$. The presence of a cross point at $n_T=0.90$ is evident, emphasizing the existence of two regimes
mediated by the intensity of the antiferromagnetic correlations. The increase in $n_T$ causes a suppression in $k_BT_C$, particularly pronounced for a more intense correlation regime. Notably, for $U = 16.0|t_0|$, the decay of $k_BT_C$ is more significant, emphasizing the connection between CDW and the pseudogap, i.e., for a large $U$ and $n_T$, the short-range antiferromagnetic correlations favor the emergence of a pseudogap and destabilize the CDW phase.

\begin{figure}[!h]
    \begin{center}
    \leavevmode
    \includegraphics[width=0.8\linewidth]{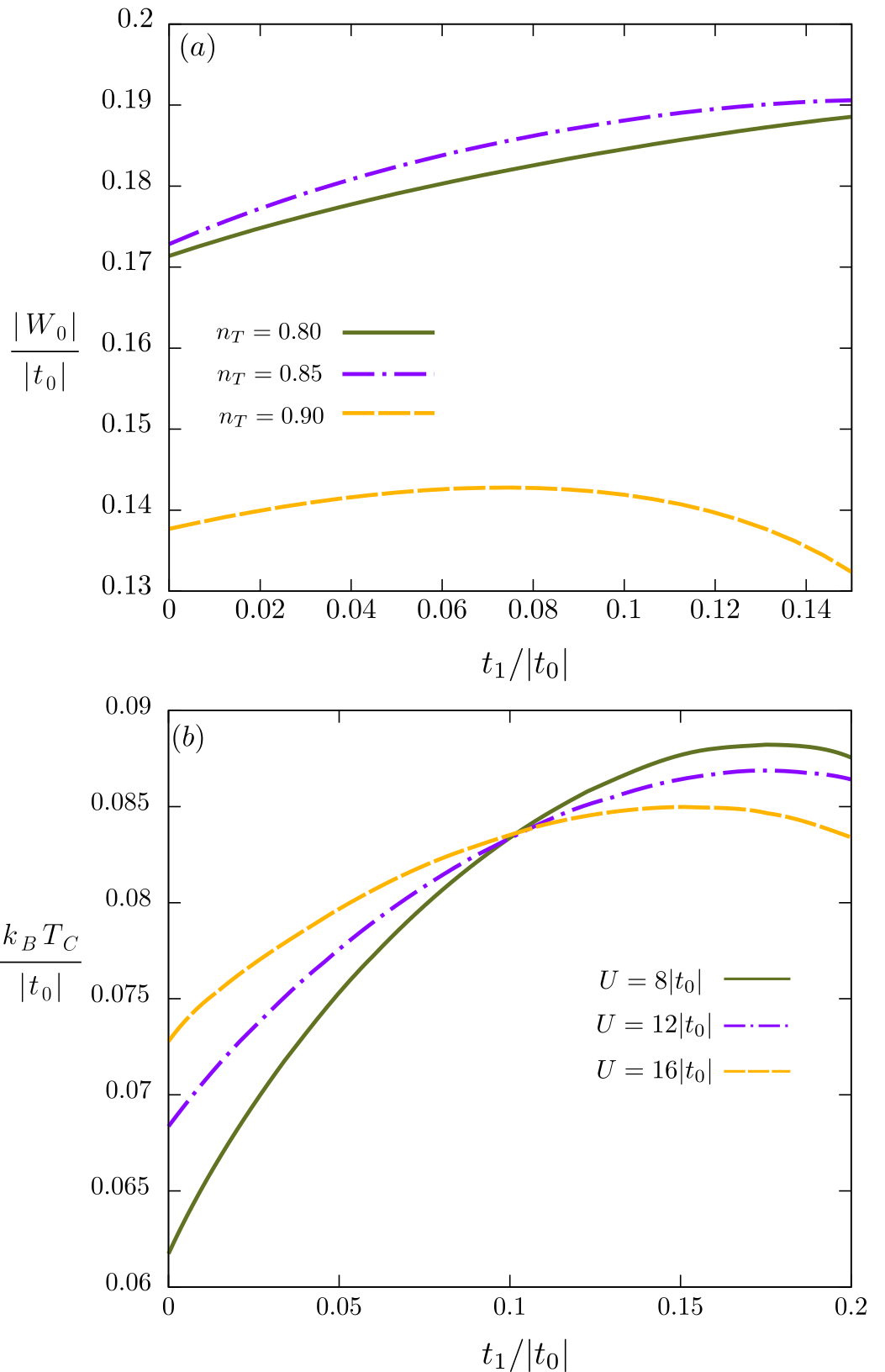}
    \end{center}
    \caption{(a) The amplitude of the CDW order parameter for $k_BT = 0.05|t_0|$, $U = 12|t_0|$ and different occupation numbers $n_T $. (b) $k_BT_C$ as a function of  $t_1/|t_0|$ for $n_T = 0.90$ and several values of $U$.}
    \label{fig:r17}
\end{figure}

As seen earlier, the next nearest neighbor hopping $t_1$ can also lead to the opening of a pseudogap (see Fig. \ref{fig:r10}). To analyze its effect on CDW, the order parameter is calculated as a function of $t_1$ for different values of $n_T$, as shown in Fig. \ref{fig:r17}(a). For $n_T = 0.80$ and $n_T = 0.85$, there is a steady increase in the order parameter amplitude, associated with the displacement of the van Hove singularity to the region near $\mu$ (see Fig. \ref{fig:r11}), favoring the CDW order.
For $n_T = 0.90$, when $t_1/|t_0| \gtrapprox 0.07$, the short-range antiferromagnetic correlations become strong enough to open a pseudogap in the normal state causing a suppression of the CDW phase.
Finally, $k_BT_C$ as a function of $t_1/|t_0|$ is shown in Fig. \ref{fig:r17}(b). Similar to the result shown in Fig. \ref{fig:r13}(b), there is  a cross point at
$\frac{t_1}{|t_0|}=0.1$.
For $\frac{t_1}{|t_0|}<0.1$, a higher $k_BT_C$ is verified for $U = 16.0|t_0|$, while for $\frac{t_1}{|t_0|}>0.1$, a higher $k_BT_C$ is related to the lower value of $U$. This result shows that  although both $U$ and  $t_1$ may enhance the antiferromagnetic correlations, an intermediate regime of antiferromagnetic correlations is necessary to stabilize the CDW phase. For sufficiently strong antiferromagnetic correlations, a pseudogap emerges and suppresses the CDW phase.

\section{Conclusions}
\label{conclusions}
\par We employed a BCS-like mean field approximation, within Green's function equation of motion formalism to describe CDW instability. To take correlations into account, the normal state uncorrelated Green's functions was replaced by a correlated one, obtained through n-pole approximation applied to the one-band Hubbard model. This was done considering the next-nearest neighbor hopping $t_1$, that along $n_T$ and $U$, act on the spin-spin correlation function $\langle \vec{S}_i \cdot \vec{S}_j\rangle$, strongly affecting the band shift $Y_{\vec{k},\sigma}$. We emphasize that $Y_{\vec{k},\sigma}$ acquires a structure in the reciprocal space that enhances such correlations. Remarkably, the effect becomes relevant at antinodal point $(\pi,0)$.  It is important to mention that the correlation functions in the band shift generally do not have an explicit dependence on the reciprocal space \cite{Roth,beenen}. However, this dependence is important \cite{calegari2011} because it displaces the bands at the nodal point $(\pi,\pi)$ and the antinodal point $(\pi,0)$ to lower energy regions. As a consequence, if $t_1$ is sufficiently large, the entire region around $(\pi,0)$ is distorted enabling the opening of a pseudogap.

Numerical results for the spectral function  $A(\vec{k},\omega=0)$ allowed the analysis of the Fermi surface in both the normal and CDW phases. For the normal state, it was shown that the next nearest neighbor hopping $t_1$ may induce a Lifshitz transition followed by the opening of a pseudogap at the antinodal points $(\pi,0)$ and $(0,\pi)$. This result is in agreement with  previous reports \cite{Chen2012,Braganca2018}, for the one band Hubbard model. Here, we argue that 
the increasing of $t_1$ enhances the short-range antiferromagnetic correlations which triggers the Lifshitz transition followed by the emergence of a pseudogap.
For the CDW phase, the Fermi surface is reconstructed due to the presence of short-range antiferromagnetic correlations associated with the spin-spin correlation function $\langle \vec{S}_i \cdot \vec{S}_j\rangle$. Indeed, these correlations distort the quasiparticle bands allowing one of the quasiparticle bands to cross the chemical potential twice at the directions $(\pi,0)$-$(0,0)$ and $(\pi,0)$-$(\pi,\pi)$, giving rise to a main Fermi surface at the CDW phase. Moreover, the main Fermi surface changes its topology from electron-like to hole-like when the total occupation $n_T$ decreases, indicating a Lifshitz transition. 

The behavior of the CDW order parameter indicates that there is an intermediate correlation regime which stabilizes the CDW phase. At the stronger correlated regime, the short-range antiferromagnetic correlations become strong and suppress the CDW order, allowing the emergence of a pseudogap.

To summarize,
we have proposed a theory based on short-range antiferromagnetic correlations to describe the connection between CDW and the pseudogap phenomenon. We have explored the connections between the pseudogap and Fermi surface reconstruction within the CDW phase through the band distortions caused by such correlations. The results show that the short-range antiferromagnetic correlations play an important role as they are responsible for a rich Fermi surface topology in both the normal and CDW phase. In the normal state, they  give rise to a pseudogap proceeded by a Lifshitz transition, while in the CDW phase, the short-range antiferromagnetic correlations are related to a reconstruction of the Fermi surface besides a Lifshitz transition.

\section*{Declaration of competing interest} 

The authors declare that they have no known competing financial
interests or personal relationships that could have appeared to influence
the work reported in this paper.

\section*{Data availability} 

Data will be made available on request.

\section*{Acknowledgments}
We would like to thank P. S. Riseborough for useful conversations.
This work was partially supported by Coordena\c{c}\~ao de Aper\-fei\c{c}oa\-mento de Pessoal de N\'{\i}vel Superior (CAPES); Conselho Nacional de Desenvolvimento Cient\'ifico
e Tecno\-l\'o\-gico  (CNPq) and
Funda\c{c}\~ao de Amparo \`a Pesquisa do Estado do Rio Grande do Sul (FAPERGS). L. Prauchner and S. G Magalhaes thank the CNPq (Conselho Nacional de Desenvolvimento Científico e Tecnológico), grant: 200778/2022-6. J. Faúndez acknowledges support from ANID Fondecyt Regular 3240320. Powered@NLHPC: This research was partially supported by the supercomputing infrastructure of the NLHPC (ECM-02).



%
\end{document}